\documentclass[twocolumn,showpacs,preprintnumbers,amsmath,amssymb,floatfix]{revtex4}
\usepackage{graphicx}

\begin{document}

\title{Strong magnetic scattering from TiO$_{x}$ adhesion layers}

\author{A. Trionfi, S. Lee, and D. Natelson}

\affiliation{Department of Physics and Astronomy, Rice University,
6100 Main St., Houston, TX 77005}

\date{\today}


\begin{abstract}
Electronic phase coherence in normal metals is incredibly
sensitive to magnetic scattering.  As a result, the weak localization
magnetoresistance and time-dependent universal conductance
fluctuations are powerful probes of magnetic impurities. We
report measurements of these effects in Au and Ag nanowires
with a 1.5 nm thick Ti adhesion layer underneath the deposited metal.  The
results indicate an anomalously large magnetic impurity
concentration due to the Ti layer.  Results suggest that this
magnetic scattering and its evolution are related to the oxidation 
state of the Ti.
\end{abstract}

\maketitle


Electronic phase coherence in normal metals gives rise to many
conductance effects in mesoscopic architectures.  These effects
include the weak localization (WL) magnetoresistance, time
dependent (TD) and magnetic field dependent (MF) universal
conductance fluctuations (UCF), and the Aharonov-Bohm effect.  This
coherence is typically limited by three types of
interactions, electron-electron scattering, electron-phonon
scattering, and spin-flip scattering from magnetic impurities.  At
temperatures much below $\sim$10 K, the electron-phonon scattering
is usually negligible.  It was suggested as early as
1984~\cite{BergmannPhysRep84} that the WL magnetoresistance could
be used as a probe of magnetic scattering at such temperatures. It
has been shown experimentally that magnetic impurity
concentrations down to a few parts per million cause observable
changes in many of the quantum transport
phenomena~\cite{BergmannJMMM86,PierrePRL02}.  Clearly, quantum transport phenomena are incredibly sensitive to even trace
quantities of magnetic impurities.

There have also been recent reports that ordinarily nonmagnetic oxides
such as HfO$_{2}$\cite{VenkatesanNature04} and TiO$_{2}$
\cite{YoonJPCM} can exhibit strong magnetic properties when oxygen
defects are present.  In this paper we report WL and TDUCF
measurements on Au and Ag nanowires that demonstrate anomalous
magnetic properties of the TiO$_{x}$ adhesion layer and strongly indicate
that the oxidation state of the Ti is responsible.  These results
again show quantum transport phenomena
as a quantitative  probe of magnetic impurities.

All the samples were patterned on undoped GaAs substrates
by standard electron beam lithography.  All metal depositions were
performed in a cryopumped electron beam evaporator that has never been
used to evaporate magnetic materials.  The evaporations
were all done as near a chamber pressure of $7.5\times10^{-7}$ Torr as
possible.  The Ti layer (when present) was deposited at a controlled
rate $\le 0.15$~nm/s, with the total thickness as detected by quartz
crystal monitor kept between 1-1.5 nm.  The Au or Ag
evaporation was then performed immediately (within minutes) without
ever removing the sample from the vacuum chamber of the evaporator.
The Au was 99.9999 \% pure while the silver was 99.99 \% pure.
Because of the affinity of Ti for oxygen, the adhesion layer is 
TiO$_{x}$, with the oxygen fraction determined in part by the
Ti deposition rate.  Without the adhesion layer, neither Au or
Ag samples exhibit signatures of magnetic scattering, as discussed
below. 

Once samples were fabricated, they were placed in a $^{4}$He
cryostat and cooled within an hour.  Once cooled to 2~K, the WL
magnetoresistance measurements were performed.  These measurements
were made using the standard ac four terminal resistance
technique.  The magnetoresistance data was acquired by changing
the applied field in steps between -.39 and .39~T in Au and
between -.33 and .33~T in Ag.  The resultant magnetoresistance
curves were analyzed using~\cite{PierrePRB03}:
\begin{eqnarray}
\frac{\Delta R}{R}\big{\vert}_{\rm 1d}& = & - \frac{e^{2}}{2 \pi
\hbar}\frac{R}{L}\left[3\left(\frac{1}{L_{\phi}^{2}}+\frac{4}{3
L_{\rm SO}^{2}}+\frac{1}{12}\left(\frac{w}{L_{\rm
B}^{2}}\right)^{2}\right)^{-1/2} \right. \nonumber\\
& & \left.  -
\left(\frac{1}{L_{\phi}^{2}}+\frac{1}{12}\left(\frac{w}{L_{\rm
B}^{2}}\right)^{2}\right)^{-1/2}\right] \label{eq:1dwlso}
\end{eqnarray}
The value $\Delta R/R$ in this equation is defined as
$R(B)-R(B=\infty)/ R(B=\infty)$ while $L_{\rm SO}$ is the spin-orbit
scattering length, $w$ is the sample width, and $L_{\rm B}\equiv
\sqrt{\hbar/2eB}$. Both the coherence length, $L_{\phi}$, and the
spin-orbit scattering length, $L_{\rm SO}$, are left as free
parameters while fitting.  The typical $L_{\rm SO}$ value was
$\sim$290 nm in Ag and $\sim$10 nm in Au.

\begin{figure}[h]
\begin{center}
\includegraphics[clip, width=7.5cm]{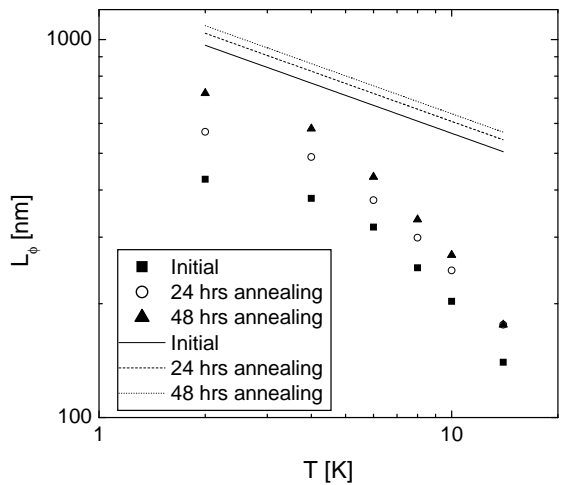}
\end{center}
\caption{Coherence lengths of a 70~nm wide Au nanowire with a 1.5~nm
thick Ti adhesion layer.  The coherence lengths were measured once
within an hour of fabrication and then after annealing in 5~Torr for
24 hours and again after an additional 24 hours. The lines are the
predicted Nyquist scattering length based on the sample
characteristics.} \label{fig:fig1}
\end{figure}

Samples cooled immediately after fabrication
systematically showed smaller coherence lengths and larger magnetic
impurity concentrations.  In one test, a 70 nm Au sample
was made with a 1.5~nm thick Ti adhesion layer and cooled to 2~K
within an hour of finishing the fabrication process.  The WL
magnetoresistance was measured, and then the sample was warmed to
305~K and left in the cryostat under low vacuum (5~Torr of helium with
some air contamination).  The sample was then cooled and the WL
remeasured.  This whole process was then repeated one more time.  The
results are shown in Figure~\ref{fig:fig1}.  The lines show the
Nyquist scattering length\cite{PierrePRB03}, the expected $L_{\phi}$
if decoherence is dominated by electron-electron scattering.

\begin{figure}[h]
\begin{center}
\includegraphics[clip, width=7.5cm]{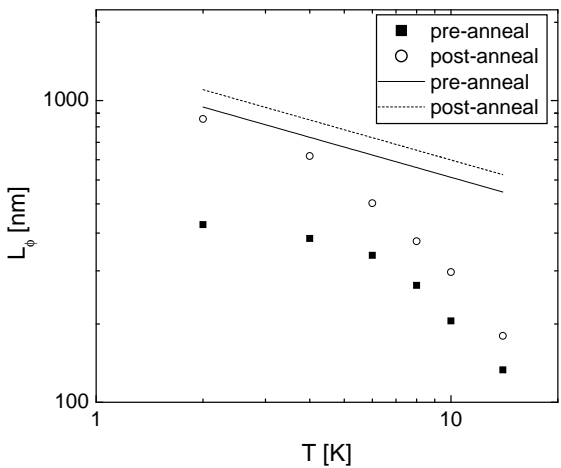}
\end{center}
\caption{Coherence lengths of a 70 nm wide Au nanowire with a 1.5
nm thick Ti adhesion layer.  The coherence lengths were measured
once within an hour of fabrication and then a second time after
letting the sample sit in ambient lab conditions for 2 weeks. The
lines are the predicted Nyquist scattering length based on the
sample characteristics.} \label{fig:fig2}
\end{figure}

Next, a similar test was performed on a 75~nm wide Au sample with a
1.5~nm thick Ti underlayer to observe how ambient lab conditions
affected the WL suppression.  The sample was placed under vacuum and
cooled to 2~K within an hour of completing fabrication.  After the WL
measurements were made, the sample was warmed and removed from the
cryostat and left to sit for roughly 2 weeks.  The sample was then
returned to the cryostat, cooled, and the WL measurements were
repeated.  The results of this test are shown in
Figure~\ref{fig:fig2}.

Samples made from Ag exhibited qualitatively the same suppression of
$L_{\phi}$ at low temperatures when a Ti adhesion layer was present,
and no suppression without the Ti layer.  Similar annealing tests were
attempted with Ag, but the oxidation of the Ag films themselves made
the test impossible over the same time scale.  The coherence lengths
of Ag with Ti did not show any noticeable change after 24 hours of air
exposure.

The Au samples clearly show a reduction in the total dephasing rate
after exposure to air for an extended period.  In light of
Ref.~\cite{YoonJPCM}, we show below that the TiO$_{x}$ layer's
approach to TiO$_{2}$ stoichiometry is likely the culprit.  The change
in the base resistance of the sample due to annealing was only $\sim$
-10\%, far too small to explain the observed change in $L_{\phi}$.
Identically prepared Au samples lacking Ti adhesion layers exhibit
similar annealing of the overall resistance, but with no sign of
suppressed $L_{\phi}$ before or after annealing.  At higher
temperatures where electron-phonon scattering dominates dephasing,
little change is observed in the samples with Ti layers.  A reduction
in the spin-flip scattering is left as the only likely cause of the
coherence length increase upon annealing.

\begin{figure}[h]
\begin{center}
\includegraphics[clip, width=7.5cm]{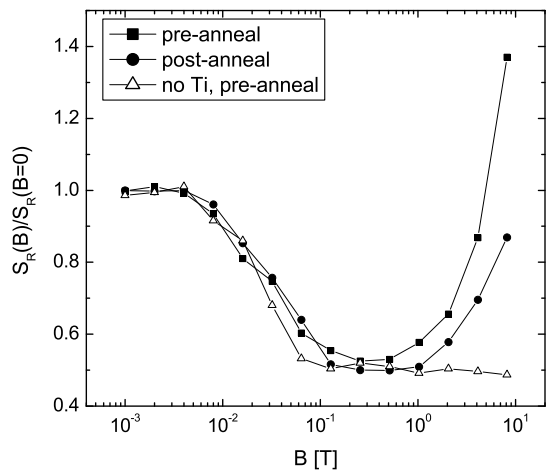}
\end{center}
\caption{The resistive noise power of two Au samples normalized to the zero
field noise power.  The closed shapes are data from a 70~nm wide
sample with a Ti layer, before and after 48 hours of annealing.  Open symbols are data from a 75~nm
wide sample with no Ti layer.} \label{fig:fig3}
\end{figure}

The TDUCF measurements provide strong evidence that the suppressed
$L_{\phi}$ prior to annealing results from magnetic scattering.  TDUCF
are an enhanced $1/f$ noise at low temperatures\cite{FengPRL86} due to
quantum interference and mobile scatterers.  Much like WL
magnetoresistance, the $1/f$ noise amplitude is altered in the
presence of a perpendicular magnetic field.  Over some field scale
related to $L_{\phi}$, the noise power will drop by a factor of
2~\cite{StonePRB89}.  A second field-dependent change occurs in the
presence of magnetic impurities.  At sufficiently large fields, Zeeman
splitting of the impurities suppresses the spin-flip
scattering~\cite{MoonPRB97}, leading to an increase in measured noise
power.  This upturn at high fields is observed in the samples with Ti,
and is reduced upon the annealing described above, while no such
upturn is seen in samples without an adhesion layer.  This is
demonstrated in Figure~\ref{fig:fig3}.

The $L_{\phi}$ data before annealing allow a quantitative examination
of the magnetic impurity concentration.  For the initial data in
Fig.~\ref{fig:fig2}, $L_{\phi}(T)$ looks nearly saturated at 2 K.
This saturation should continue until the temperature is well below
the Kondo temperature of the magnetic impurities.  The Nagaoka-Suhl
formula relating the spin-flip scattering rate to the concentration of
magnetic impurities can be used to estimate the this impurity
concentration.  By assuming the saturated $L_{\phi}$ is dominated by
spin-flip scattering, the relation should be~\cite{PierrePRB03}:
\begin{equation}
\tau_{\phi}^{\textrm{saturated}} [\textrm{ns}] \simeq
.6/c_{\textrm{mag}} [\textrm{ppm}] \label{eq:estmagimp}
\end{equation}
Here coherence time is related to $L_{\phi}$ via the classical
diffusion constant, $L_{\phi}\equiv\sqrt{D\tau_{\phi}}$.  In this
sample this gives an effective bulk magnetic impurity concentration of
$\sim$16 ppm, much higher than the expected total impurity
concentration of the source gold (1 ppm) or Ti (8~ppm).  Assuming
spin-flip scatterers reside at the TiO$_{x}$/Au interface, we can
estimate the surface density of impurities at 14000
impurities/$\mu$m$^{2}$.  Attempts to examine the direct magnetic
signal from these moments via SQUID magnetometry and low temperature
electron spin resonance measurements were unsuccessful.

To further confirm the important role of oxygen, an examination of the
coherence length suppression as a function of the Ti evaporation rate
was performed.  With the chamber vacuum near $7.5\times10^{-7}$ Torr
when the Ti evaporation begins, much of the Ti adhesion layer is
actually TiO$_{x}$ by the time the Ag or Au is deposited over it.
Lower evaporation rates of the Ti should yield layers with higher
oxygen content, closer to TiO$_{2}$, and presumably with a decreased
magnetic response.  Table~\ref{tab:tab1} shows the WL
magnetoresistance suppression at 2~K as a function of the Ti
evaporation rate.  Since both $L_{\phi}(T)$ and the predicted Nyquist
time depend on the diffusion constant of the metal, properly comparing
different samples must be done with care.  We compared the ratio of
the measured $\tau_{\phi}$ to the expected Nyquist scattering time for
that particular sample.  The clear correlation between the Ti
evaporation rate and the coherence suppression supports the
hypothesis that the oxygen content of the Ti layer strongly affects
that layer's magnetic scattering properties.

\begin{table}
\caption{Au and Ag samples prepared with a Ti adhesion layer
deposited at different rates.  The ratio of the measured coherence
time to the expected Nyquist scattering time is shown at 2 K.}
\begin{tabular}{c c c}
\hline \hline
Material & Ti rate [\AA/s] & $\tau_{\phi}/\tau_{\textrm{e-e}}$ \\
\hline
Ag & - & .82 \\
Ag & .5 & .36 \\
Ag & 1.5 & .14 \\
Au & - & 1.14 \\
Au & .3 & .71 \\
Au & 1 & .19 \\
\hline \hline
\end{tabular}
\label{tab:tab1}
\end{table}

The role of metallic Ti as a magnetic scatterer has been examined
previously\cite{BergmannPRB95} in the limit of extremely dilute
coverage, with an observed suppression of the weak localization
magnetoresistance.  The proposed mechanism in that case was local spin
fluctuations, with a dephasing rate proportional to $T$.  Such a
temperature dependence is not compatible with our observations of a
weakening dependence of $L_{\phi}$ on temperature as $T \rightarrow
0$.

In summary, measurements of quantum corrections to the conduction in
Au and Ag nanowires prepared with adhesion layers confirm strong
magnetic effects in nonstoichiometric TiO$_{x}$.  Although standard
SQUID and ESR measurements failed to detect a magnetic signature in
Ti/Au films, it may be possible to detect a signal with scanning SQUID
microscopy~\cite{KirtleyRSI01}.  Such studies would give important
information about the microscopic nature of magnetic defect centers in
ordinarily nonmagnetic oxides.

The authors thank N.O. Birge and G. Bergmann for helpful discussions.  
This work was supported by the David and Lucille Packard
Foundation and DOE Grant No. DE-FG03-01ER45946/A001.




\end{document}